\documentclass[APS,PRB,twocolumn,showpacs]{revtex4}
\usepackage{amssymb}
\usepackage{txfonts}
\usepackage{graphicx}

\begin{document}

\title[]{Upper critical field, anisotropy, and superconducting properties of Ba$_{1-x}$K$_x$Fe$_2$As$_2$ single crystals}

\author{Zhao-Sheng Wang, Hui-Qian Luo, Cong Ren and Hai-Hu Wen}\email{hhwen@aphy.iphy.ac.cn}

\address{National Laboratory for Superconductivity, Institute of Physics and Beijing National Laboratory for Condensed Matter Physics, Chinese Academy of
Sciences, P.O.Box 603, Beijing 100190, China}

\begin{abstract}

The temperature dependent resistivity of Ba$_{1-x}$K$_x$Fe$_2$As$_2$
(x = 0.23, 0.25, 0.28 and 0.4) single crystals and the angle
dependent resistivity of superconducting
Ba$_{0.6}$K$_{0.4}$Fe$_2$As$_2$ single crystals were measured in
magnetic fields up to 9 T. The data measured on samples with
different doping levels revealed very high upper critical fields
which increase with the transition temperature, and a very low
superconducting anisotropy ratio $\Gamma=H_{c2}^{ab}/H_{c2}^c
\approx$ 2. By scaling the resistivity within the framework of the
anisotropic Ginzburg-Landau theory, the angle dependent resistivity
of the Ba$_{0.6}$K$_{0.4}$Fe$_2$As$_2$ single crystal measured with
different magnetic fields at a certain temperature collapsed onto
one curve. As the only scaling parameter, the anisotropy $\Gamma$
was alternatively determined for each temperature and was found to
be between 2 and 3.

\end{abstract}

\maketitle

Since the discovery of superconductivity at 26 K in the FeAs-based
new superconductor LaFeAsO$_{1-x}$F$_x$\cite{sample1}, great
interests have been stimulated in the community of condensed matter
physics and material sciences. Under intensive study, the
superconducting transition temperature ($T_c$) was quickly promoted
to 55 K by replacing La with Sm\cite{55KRen}, making the iron-based
superconductors to be non copper-based materials with $T_c$
exceeding 50 K. At the same time, hole-doped superconductors were
also synthesized successfully in La$_{1-x}$Sr$_x$FeAsO
\cite{ourhole} and Ba$_{1-x}$K$_x$Fe$_2$As$_2$\cite{Rotter} systems.
Beside the high critical temperature, the upper critical field was
found to be very high in the iron-based
superconductors\cite{zhu,high-field,jiaying,Altarawneh,Yuan}. Many
theoretical models have been proposed to give explanations for the
mechanism of superconductivity, such as non-phonon pairing
mechanisms\cite{non-phonon}, multi-band
superconductivity\cite{multi-band}, etc.. Meanwhile, many
experiments have revealed evidence for an unconventional paring
mechanism in the single layer structure iron-based superconductors,
such as point contact tunneling spectroscopy\cite{shanlei},
NMR\cite{Buechner,zhengguoqing}, specific heat\cite{mugang}, lower
critical field\cite{CR,30}, etc.. However, most of these
measurements were carried on polycrystalline samples because of the
lack of sizable single crystals in the system with a single layer
structure.

Fortunately, the Ba$_{1-x}$K$_x$Fe$_2$As$_2$ single crystals can be
grown by metal or self-flux method. With this success, more accurate
measurements have become possible in this iron-based superconducting
system. As one of the basic parameters, the superconducting
anisotropy $\Gamma=H_{c2}^{ab}/H_{c2}^c$ is crucial for both
understanding the superconducting mechanism and the potential
applications, where $H_{c2}^{ab}$ and $H_{c2}^c$ are the upper
critical fields when the magnetic field is applied within the
ab-plane and c-axis, respectively. With a layered structure in the
FeAs-based superconductors, such as cuprates, strong anisotropy of
superconductivity might be expected.\cite{Singh} An estimation of
$\Gamma \geq$ 30 was made on (Nd, Sm)FeAsO$_{0.82}$F$_{0.18}$
polycrystals from the $c-$axis infrared plasma frequency\cite{30}.
However, an anisotropy of about 4-6 for the upper critical field was
found in NdFeAsO$_{0.82}$F$_{0.18}$ single crystals\cite{jiaying}
based on the transport measurements. Therefore, it is very necessary
to determine the upper critical fields and the superconductivity
anisotropy of the Ba$_{1-x}$K$_x$Fe$_2$As$_2$ system, especially for
different doping levels. In this work, we present the temperature
dependent resistivity of Ba$_{1-x}$K$_x$Fe$_2$As$_2$ (x = 0.23,
0.25, 0.28, and 0.4) single crystals and the angle dependent
resistivity of Ba$_{0.6}$K$_{0.4}$Fe$_2$As$_2$ single crystals. A
very high upper critical field and a rather low superconductivity
anisotropy were found. Both the upper critical field and the
anisotropy $\Gamma$ have been obtained for samples with different
doping levels of potassium at different temperatures.

\begin{figure}
\includegraphics[scale=0.7]{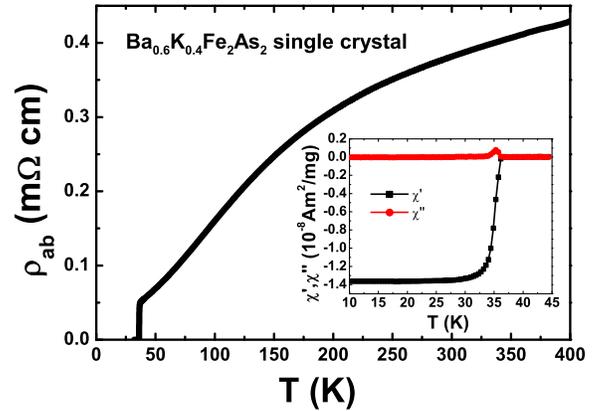}
\caption{(color online) Temperature dependence of the in-plane
electrical resistivity for a Ba$_{0.6}$K$_{0.4}$Fe$_2$As$_2$ single
crystal in zero field up to 400 K. One can see that the
$\rho_{ab}(T)$ curve exhibits a continued curvature up to 400 K. The
inset shows the temperature dependent AC susceptibility for the
Ba$_{0.6}$K$_{0.4}$Fe$_2$As$_2$ single crystal. Sharp
superconducting transitions are obvious both in the resistive and
the AC susceptibility data.}
\end{figure}

The single crystals of Ba$_{1-x}$K$_x$Fe$_2$As$_2$ (0 $< x \leq$
0.4) were grown by using FeAs as the self flux at ambient pressures.
During synthesis the FeAs content is several times higher than that
of $Ba_{1-x}K_x$, and the actual composition of potassium is
controlled by the growing temperature and the relative composition
ratio between Ba and K. Details for the synthesis were described in
Ref.\cite{ours}. The composition of potassium given in this Rapid
Communication is obtained from the energy dispersive x-ray (EDX)
analysis spectrum. The typical lateral sizes of the crystals are
about 2 to 3 mm, while thickness is about 0.1 mm. The good $c$-axis
orientation of the crystals has been demonstrated by the x-ray
diffraction (XRD) analysis which shows only the sharp ($00\ell$)
peaks. The temperature dependent ac susceptibility of the crystals
were measured on an Oxford cryogenic system Maglab-EXA-12. During
the measurement, an alternating magnetic field (H = 1 Oe)
perpendicular to the $ab$-planes of the crystal at a frequency f=333
Hz was applied. The susceptibility curve of a
Ba$_{0.6}$K$_{0.4}$Fe$_2$As$_2$ single crystal is shown in the inset
of Figure 1. The superconducting transition is very sharp with
$\Delta T_c < 0.5$ K (1\%-90\%$\rho_n$, where $\rho_n$\ is the
normal state resistivity), demonstrating the high quality and
homogeneity of the single crystal.

\begin{figure}
\includegraphics[scale=0.95]{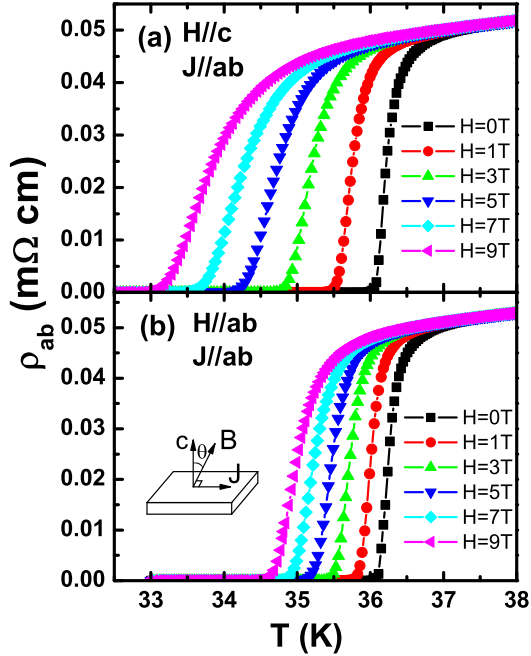}
\caption{(color online) Temperature dependence of the in-plane
electrical resistivity for the Ba$_{0.6}$K$_{0.4}$Fe$_2$As$_2$
single crystal at fields $\mu_0H$ = 0, 1, 3, 5, 7, 9 T with (a) $H
\parallel c$ and (b)$H \parallel ab$, respectively. The inset of (b)
illustrates the definition of angle $\theta$.}
\end{figure}

\begin{figure}
\includegraphics[scale=0.85]{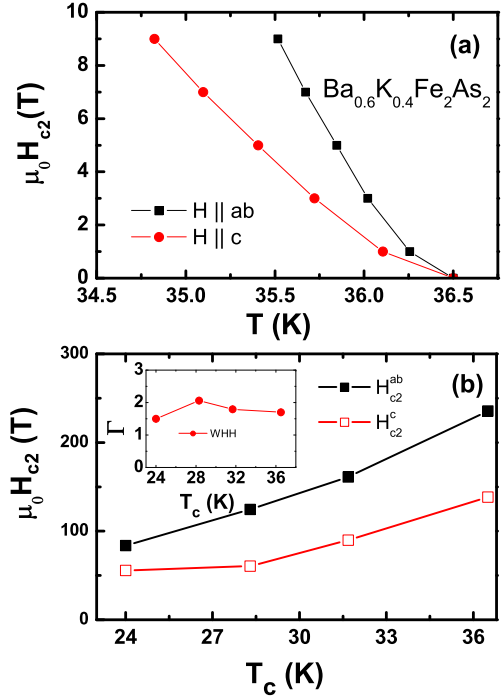}
\caption{(color online) (a) The upper critical field of
Ba$_{0.6}$K$_{0.4}$Fe$_2$As$_2$ single crystal for H $\parallel$ c
and H $\parallel$ ab respectively. (b) The upper critical field
versus $T_c$ of Ba$_{1-x}$K$_x$Fe$_2$As$_2$ crystals (x = 0.23,
0.25, 0.28, and 0.4) single crystals with $T_c$ = 24, 28.3, 31.7,
36.5 K. The inset of (b) presents the anisotropy ratio of
$H_{c2}(0)$ along c-axis and ab-planes for the four samples. The
lines are guides to eyes.}
\end{figure}

The resistivity measurements were carried out on a physical property
measurement system (PPMS)(Quantum Design) with magnetic fields up to
9 T.  The electrical resistivity of Ba$_{1-x}$K$_x$Fe$_2$As$_2$ ($x$
= 0.23, 0.25, 0.28, and 0.4) single crystals and the angle dependent
resistivity of Ba$_{0.6}$K$_{0.4}$Fe$_2$As$_2$ single crystal were
measured by the standard four-probe method. Figure 1 presents the
in-plane resistivity $\rho_{ab}$ of the
Ba$_{0.6}$K$_{0.4}$Fe$_2$As$_2$ single crystal in zero field up to
400 K. One can see that the $\rho(T)$ data exhibit a continued
curvature up to 400 K. In the angle-resolved resistivity
measurement, the angle $\theta$ was varied from $0^{\circ}$ to
$180^{\circ}$, where $\theta = 0^{\circ}$ corresponded to the
configuration of $H
\parallel c-$axis and $\theta = 90^{\circ}$ to $H \parallel
ab-$plane, respectively. The current was applied in the $ab-$plane
and perpendicular to the magnetic field in all cases (as shown in
the inset of Figure 2(b)).  The sample with x = 0.4 exhibited a
sharp resistive superconducting transition at $T_c\simeq 36.5$ K
(90\% of the normal state resistivity) with $\Delta T_c < 0.5$ K.
The residual resistivity is about $\rho(38K)=0.05 m\Omega cm$ and
residual resistivity ratio is found to be
$RRR=\rho(300K)/\rho(38K)=$7.4.

Many experiments had revealed high upper critical fields in the
LnFeAsO$_{1-x}$F$_x$(The estimated $H_{c2}^{ab}$ is beyond 50 T for
Ln = La, and beyond 100 T for Ln = Sm, Pr, and Nd)
system\cite{high-field,zhu,jiaying}. In the
Ba$_{1-x}$K$_x$Fe$_2$As$_2$ system, the value of the upper critical
field seems also very high.\cite{Altarawneh,Yuan} Figure 2 (a) and
(b) show the temperature dependent resistivity $\rho(T)$ curves of
the Ba$_{0.6}$K$_{0.4}$Fe$_2$As$_2$ single crystal at magnetic
fields up to 9 T along the $c$-axis and $ab$ planes, respectively.
It is found that the superconducting transitions are broadened
slightly, which indicates that the upper critical field should be
very high. One can also see that the resistive transition curves
shift parallel down to lower temperatures upon using a magnetic
field, this may suggest a field induced pair breaking effect in the
present system. This is again very different from the case in the
cuprate superconductors where the $\rho(T)$ broadens by exhibiting a
fan structure with the onset transition part barely changed by the
magnetic field. From the transition curves in Fig.2(a) and Fig.2(b),
we have already an idea that the anisotropy ratio is quite small. In
Figure 3(a) we present the $H_{c2}$-T curves for the
Ba$_{0.6}$K$_{0.4}$Fe$_2$As$_2$ single crystal for both $H
\parallel c$ and $H
\parallel ab$, respectively. The H$_{c2}(T)$ is determined at the point where  $\rho=90\%\rho_n$. The curves of $H_{c2}(T)$ are very
steep with average slopes $-dH^{ab}_{c2}/dT_c|_{T_c}$ = 9.35 T/K for
$H
\parallel ab$ and $-dH^{c}_{c2}/dT_c|_{T_c}$ = 5.49 T/K for $H
\parallel c$. According to the Werthamer-Helfand-Hohenberg (WHH)
formula\cite{WHH} $H_{c2}=-0.69(dH_{c2}/dT)|_{T_c}T_c$ and taking
$T_c$ = 36.5 K, the values of upper critical fields are
$H^{ab}_{c2}(0)$ = 235 T and $H^{c}_{c2}(0)$ = 138 T. Although these
high values of upper critical fields may subject to a modification
when the direct measurements are done in the high field experiments,
we believe that the FeAs-based superconductors are really robust
against the magnetic field. For example, taking the zero temperature
values we obtained using the WHH formula, we have
$\mu_0H^{ab}_{c2}(0)/k_BT_c$ = 235/36.5 T /K = 6.43 T/K. This ratio
is much beyond the Pauli limit $\mu_0H_{c2}(0)/k_BT_c$ = 1.84 T/K
for a singlet pairing when the spin-orbital coupling is
weak\cite{Tinkham}. This may manifest an unconventional mechanism of
superconductivity in this material. The values of $H_{c2}(0)$ for
other three samples with different doping levels of potassium and
thus transition temperatures were also determined in the same way.
The results are shown in Figure 3 (b). It is found that the
$H_{c2}(0)$ decreases quickly with the decrease of $T_c$.  We note
that a recent result reported that the WHH approximation may not be
simply applied in these materials\cite{not WHH}, and even
$H_{c2}(0)$ may be affected by the multiband property. However, our
results clearly indicate that the upper critical fields in the
present system are really very high without any doubt. The large
value of the ratio $\mu_0H^{ab}_{c2}(0)/k_BT_c$ was also found in
our previous measurements on $NdFeAsO_{0.82}F_{0.18}$ single
crystals\cite{jiaying}. As far as we know, the large value of the
ratio $\mu_0H_{c2}(0)/k_BT_c$ obtained in FeAs-based superconductors
has been rarely reported in other superconductors.

\begin{figure}
\includegraphics[scale=0.8]{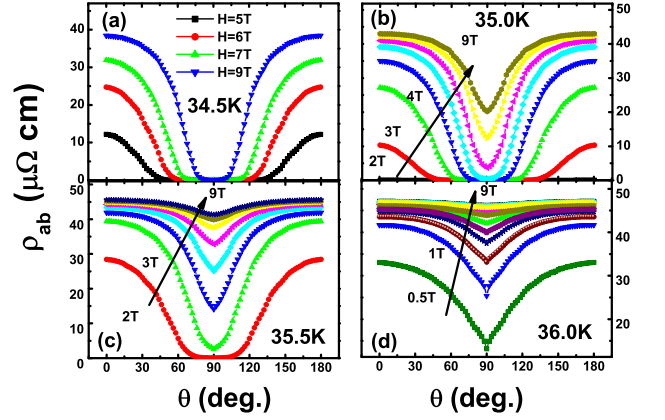}
\caption{(color online) Angular dependence of resistivity at (a)
34.5 K in $\mu_{0}H=$ 5, 6, 7, 9 T, (b) 35 K in $\mu_{0}H=$ 2, 3, 4,
5, 6, 7, 8, 9 T, (c) 35.5 K in $\mu_{0}H=$ 2, 3, 4, 5, 6, 7, 8, 9 T,
(d) 36 K in $\mu_{0}H=$ 0.5, 1, 1.5, 2, 2.5,  3, 4, 5, 6, 7, 8, 9 T
for the Ba$_{0.6}$K$_{0.4}$Fe$_2$As$_2$ single crystal.}
\end{figure}

\begin{figure}
\center
\includegraphics[scale=0.75]{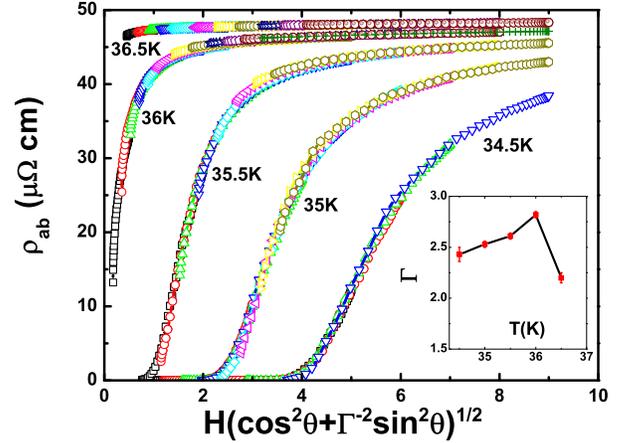}
\caption{(color online) Scaling of the resistivity versus
$\tilde{H}=H\sqrt{\cos^2(\theta)+\Gamma^{-2}\sin^2(\theta)}$ at 34,
34.5, 35, 35.5, 36, 36.5, 37 K in different magnetic fields. Each
curve is scaled nicely by adjusting $\Gamma$.  The inset presents
the temperature dependent $\Gamma(T)$ for the
Ba$_{0.6}$K$_{0.4}$Fe$_2$As$_2$ single crystal.  The line is guided
to eyes.}
\end{figure}

In an anisotropic type-II superconductor, the magnetic field
destroys superconductivity at the upper critical fields
$H^{ab}_{c2}$ and $H^{c}_{c2}$ for applied fields $H \parallel ab$
and $H\parallel c$, respectively. The effective upper critical field
varies between the two orientations depending on the superconducting
anisotropy ratio $\Gamma$ = $H^{ab}_{c2}/H^{c}_{c2}$. In the inset
of Fig.3(b), we show the doping dependence of the anisotropy ratio
$H^{ab}_{c2}(0)/H^{c}_{c2}(0)$ for the samples with different doping
levels. One can see that the anisotropy for different samples all
locate around 2. This result is surprising to us since the band
structure calculations by Singh\cite{Singh2} clearly show that the
Fermi surface sheets and dimensionality strongly depends on the
doping level. For the Ba$_{0.6}$K$_{0.4}$Fe$_2$As$_2$ single crystal
it is found that $\Gamma \approx$ 1.70 $\sim$ 1.86 at zero
temperature, and $\Gamma$ are below 2.1 for all other samples. This
value is quite close to that derived on similar samples also from
the shift of the resistive transitions under magnetic
fields.\cite{Canfield} The values of anisotropy are rather small in
comparison with all cuprate superconductors, and slightly lower than
that of F-doped NdFeAsO with $\Gamma$ = 4 -
6\cite{jiaying,jiaying2}.

The anisotropy ratio determined above may subject to a modification
because of the uncertainties in determining the upper critical field
value by taking different criterions of resistivity and in using
different formulas. One major concern was that the zero temperature
value $H_{c2}(0)$ was determined by using the experimental data near
$T_c$. This concern can be removed by the measurements of angular
dependent resistivity. According to the anisotropic Ginzburg-Landau
theory, the resistivity in the mixed state depends on the effective
field $H/H_{c2}^{GL}(\theta)$. In this case the resistivity measured
at different magnetic fields but at a fixed temperature should be
scalable with the variable $H/H_{c2}^{GL}(\theta)$. The effective
upper critical field $H_{c2}^{GL}(\theta)$ at an angle $\theta$ is
given by

\begin{equation}
H_{c2}^{GL}(\theta) =
H_{c2}^{c}/\sqrt{\cos^2(\theta)+\Gamma^{-2}\sin^2(\theta)}.
\end{equation}

Thus using the scaling variable
$\tilde{H}=H\sqrt{\cos^2(\theta)+\Gamma^{-2}\sin^2(\theta)}$, the
resistivity should collapse onto one curve in different magnetic
fields at a certain temperature\cite{Blatter} when an appropriate
$\Gamma$ value is chosen. Figure 4 presents four sets of data of
angular dependence of resistivity at 34.5, 35, 35.5, and 36 K for
the Ba$_{0.6}$K$_{0.4}$Fe$_2$As$_2$ single crystal. At each
temperature, a cup-shaped feature centered around $\theta = 90^o$ is
observed. As shown in Fig. 5, the curves measured at different
magnetic fields but at a fixed temperature are scaled nicely by
adjusting $\Gamma$. The values of $\Gamma$ were thus obtained for
temperatures 34.5, 35, 35.5, 36, and 36.5 K. Because only one
fitting parameter $\Gamma$ is employed in the scaling for each
temperature, the value of $\Gamma$ is more reliable compared with
the one determined from the ratio of $H_{c2}^{ab}$ and $H_{c2}^{c}$
as used above. But both methods yield similar values of $\Gamma$,
which implies the validity of the values determined in this work. It
is found that, the anisotropy increases from 2.43 for 34.5 K to 2.82
for 36 K, and then decreases slightly for 36.5 K, as plotted in the
inset of Figure 5. Actually the anisotropy was also measured to the
much higher field and lower temperature region\cite{Altarawneh,Yuan}
in the same system and was found to decrease with temperature and
finally reached a value of about 1-1.5 in the low T limit. This kind
of temperature dependence of $\Gamma(T)$ is not expected for a
single band anisotropic superconductor and needs a further check in
other measurements. It may be attributed to the the effect of
two-gap scenario\cite{twogap,multiband,ARPES,CR2}. In addition, it
should be noted that the good scaling behavior suggests a
field-independent anisotropy in the temperature and field range we
investigated. The small anisotropy can be qualitatively understood
based on the recent band structure calculations\cite{Singh2} in
which it is shown that the Fermi surface sheets are not 2D cylinder
like but rather exhibit a complicated 3D feature with a quite strong
dispersion along c-axis. Our results here, very high upper critical
fields and very low anisotropy, should be stimulating in fulfilling
a quantitative calculation on the electronic structure of the doped
samples, and ultimately providing an understanding to the underlying
mechanism of superconductivity.

In conclusion, we have investigated the temperature dependent
resistivity for Ba$_{1-x}$K$_x$Fe$_2$As$_2$ (0 $< x \leq$ 0.4)
single crystals in magnetic fields up to 9 T. It is found that the
system poses a very high upper critical field and a very low
superconducting anisotropy ratio which is around 2 for all the
samples. In an alternative way, we also determined the anisotropy
ratio by investigating the angle dependent resistivity in the
Ba$_{0.6}$K$_{0.4}$Fe$_2$As$_2$ single crystals. Both methods yield
the similar values of the anisotropy ratio $\Gamma$. Our results
strongly suggest that the anisotropic Ginzburg-Landau theory can be
used very well to describe the data in the mixed state.

This work is supported by the Natural Science Foundation of China,
the Ministry of Science and Technology of China (973 project No:
2006CB60100, 2006CB921107, 2006CB921802), and Chinese Academy of
Sciences (Project ITSNEM).

\end{document}